\documentclass[final,1p,times]{elsarticle}

\usepackage{amsmath,amssymb,amsfonts,amscd}
\usepackage{graphicx}

\usepackage{color,soul}

\begin{document}

\begin{frontmatter}

\title{Interference and inequality in quantum decision theory}
\author[label1]
{Taksu Cheon \corref{cor1}}
\ead{taksu.cheon@kochi-tech.ac.jp}
\author[label2]
{Taiki Takahashi}
\ead{ttakahashi@lynx.let.hokudai.ac.jp}
\address
[label1]
{Laboratory of Physics, Kochi University of Technology,
Tosa Yamada, Kochi 782-8502, Japan}
\address
[label2]
{Laboratory of Social Psychology, Department of Behavioral Science,
Faculty of Letters, 
Hokkaido University,\\
N.10, W.7, Kita-ku, Sapporo, 060-0810, Japan}

%
\cortext[cor1]{corresponding author}

\date{\today}

\begin{abstract}
The quantum decision theory is examined in its simplest form of 
two-condition two-choice setting.  
A set of inequalities to be satisfied by any quantum conditional probability
describing the decision process is derived. 
Experimental data indicating the breakdown of classical explanations are
critically examined with quantum theory using the full set of quantum phases.
\end{abstract}

\begin{keyword}
contextual probability \sep sure-thing principle \sep
quantum phase
\PACS 89.65.Ef \sep 03.65.+w \sep 89.65.Gh\\
%
%
\end{keyword}

\end{frontmatter}


\section{Introduction}
%
There has been growing recognition that the quantum probability, as an intriguing extension of classical probability, may find its application well beyond the microscopic realm of atoms and elementary particles.
Among them, an interesting possibility of applying quantum description on psychological 
process has attracted much recent attentions.
Several authors have  noted  \cite{AA94,PB09,KH09}  that the quantum interference 
among different paths of events leading to the same decision can account for
the paradoxical experimental observation of
the {\it violation of sure-thing principle} \cite{ST92, TS92}, that 
has mystified psychologists for quite some time.
The said paradox refers to the two-choice experiment under two preconditions, in which
people make certain choice under the first condition and also make the identical choice under the second condition, but make the opposite choice under the situation where the precondition is supposed to be an unknown combination of the first and the second. 
This experimentally observed phenomenon is in direct contradiction with the fundamental assumption of the independence axiom in classical decision making theory \cite{NM47}, that is built upon the concept of classical (Bayesian) probability \cite{SA54}.  There is now a glimmer of hope that the quantum probability might provide a basis for a unifying theory of human cognition which has been long sought-after \cite{PB09}.

%
The key concept in the quantum decision theory is the {\it interference} among the probabilities of choices in different preconditions.  This is a direct result of the central assumption of quantum decision theory that the preconditions and the choices are represented by vectors residing in a Hilbert space, and the unknown precondition is represented by linear superposition of known preconditions.
The interference term has a form of {\it geometric mean} of probabilities under known conditions that are to be added to the classical arithmetic mean, thus representing possible alternate psychological mechanism to the standard theory. 
The amount of interference is controlled by the phase parameters that are inherent
in Hilbert space vectors.  Appearance of these phase parameters, which can be adjusted to explain experimental numbers, has been at the heart of the seeming success of the quantum explanation of the violation of sure thing principle.  Unfortunately, a simple and easy-to-understand presentation of quantum decision theory based on a coherent framework seems to be still lacking.  Moreover, a systematic analysis of experimental numbers based on a single framework has been missing.
As it stands, therefore, it seems hard either to prove or disprove 
the effectiveness of quantum description of decision making process.

In this note, we reexamine the quantum analysis of decision making process 
in its simplest form, 
with the use of projection operator formalism,
to clarify the essential elements of the theory,
and identify the minimal number of phase parameters involved in the
quantum description.
We further derive a set of inequalities for conditional probabilities that involve only 
classically observable quantities, that have to be satisfied by {\it any} quantum
description, thus are suitable to  judge the utility of quantum approaches
to decision theory.
We illustrate the procedure to extract quantum phase parameters uniquely 
in systematic fashion, using the data obtained from previous psychological experiments.
Hopefully, with accumulations of more experimental data, 
our approach will  eventually enable the critical examination of quantum 
decision making theories. 

\section{Preliminaries}

As a preliminary exercise, we start by examining  a trivial, 
but usually neglected relation binding the quantum probabilities.
Consider a two level system in the state
%
$
\left| \psi \right> = 
  \psi_0 \left| 0 \right>
+\psi_1 \left| 1 \right> .
$
%
The probabilities of finding the system in
the state $\left| 0 \right>$ and $\left| 1 \right>$  are given, respectively by
%
$p_0 = \left| \left< 0 | \psi \right> \right|^2 = \psi_0^* \psi_0$
and
$p_1 = \left| \left< 1 | \psi \right> \right|^2 = \psi_1^* \psi_1 $,
%
which places the constraint $\psi_0^* \psi_0+\psi_1^* \psi_1=1$ on 
quantum amplitudes $\psi_j$.
We ask a question what the probability of finding the system in an intermediate state
%
$
\left| K \right> = \kappa_0 \left| 0 \right> + \kappa_1 \left| 1 \right>
$
%
is.  The proportion of the states $\left| 0 \right>$ and $\left| 1 \right>$ in
$\left| K \right>$ are given by $q_0 = \kappa_0^*\kappa_0$ and
$q_1 = \kappa_1^*\kappa_1$, respectively, with constraint $q_0+q_1=1$.   
The answer is immediately obtained as 
$p_K = \left| \left< K | \psi \right> \right|^2$ in the form
\begin{eqnarray}
\label{QAV}
p_K  = 
\frac{(2q_0p_0)+(2q_1p_1)}{2} + \sqrt{(2q_0p_0) (2q_1p_1)} \cos\theta  ,
\end{eqnarray}
where the relative quantum phase $\theta$ is defined through
%
$\kappa_0^* \kappa_1 \psi_0^* \psi_1 = e^{i \theta} \sqrt{q_0  q_1 p_0 p_1}$. 
%
The first term of (\ref{QAV}) is 
the arithmetic mean of joint probabilities $q_0p_0$ and $q_1p_1$, 
as expected in usual classical intuition, while the second term,
representing the quantum interference is given in the form of the geometric mean
with the weight given by cosine of the quantum phase. With the obvious relation
$-1 \le \cos \theta \le 1$, we find
\begin{eqnarray}
\label{QAINQ}
\left( \sqrt{q_0p_0}-\sqrt{q_1p_1} \right)^2 
\le p_K  \le
\left( \sqrt{q_0p_0}+\sqrt{q_1p_1} \right)^2 .
\end{eqnarray}
%

\section{Quantum conditional probabilities}
Let us consider an agent facing a choice between two
actions which we call $\left|0\right>$ and $\left|1\right>$.
Let us further assume that his choice is conditional to
an event that can take two outcome, $\left|0\right)$ and $\left|1\right)$,
that precedes his choice.
The quantum mechanical description of this agent is achieved by
the state
\begin{eqnarray}
\label{fullwf}
\left| \left| \Psi\right) \right> = 
\left| 0 \right) \left|\!\right. \psi^{(0)} \left.\!\right>
+\left| 1 \right) \left|\!\right. \psi^{(1)} \left.\!\right> ,
\end{eqnarray}
%
with
\begin{eqnarray}
\label{agwf}
&&\!\!\!\!\!\!\!\!\!\!\!\!\!\!\!\!
\left|\!\right. \psi^{(0)} \left.\!\right> =
\psi^{(0)}_0 \left| 0 \right> + \psi^{(0)}_1 \left| 1 \right> ,
\nonumber \\
&&\!\!\!\!\!\!\!\!\!\!\!\!\!\!\!\!
\left|\!\right. \psi^{(1)} \left.\!\right> =
\psi^{(1)}_0 \left| 0 \right> + \psi^{(1)}_1 \left| 1 \right> ,
\end{eqnarray}
where four $\psi^{(k)}_j$ are complex numbers.
The conditional probability of agent taking action $\left|j\right>$ 
after observing the event $\left|k\right)$ is given 
by $p^{(k)}_j = | \psi^{(k)}_j |^2$.
Naturally, $\psi^{(k)}_j$ are constrained by the normalization
\begin{eqnarray}
\psi^{(k)*}_0\psi^{(k)}_0+\psi^{(k)*}_1\psi^{(k)}_1=1
\end{eqnarray}
for $k=0, 1$.
Let us now suppose that there is an intermediate event $\left| K \right)$ described by
\begin{eqnarray}
\left| K \right) = \kappa^{(0)} \left| 0 \right)+ \kappa^{(1)} \left| 1 \right),
\end{eqnarray}
with complex numbers $\kappa^{(k)}$ satisfying the constraint
%
$
\kappa^{(0)*}\kappa^{(0)}+\kappa^{(1)*}\kappa^{(1)}=1 .
$
%
For this intermediate event,  the chance of the event $\left(k\right)$ occurring is given by
$q^{(k)} = |\kappa^{(k)}|^2 = \left(k|K\right)\left(K|k\right)$ .

Let us now consider the quantum state after the occurrence of the event 
$\left|K\right)$.
This process can be thought of as a quantum measurement.
Suppose a state $\left| \phi \right)$ is measured by an observer
who finds the system in the state $\left| K \right)$.  This process can be 
described by the application of non-unitary projection operator 
$\left| \phi \right> \to {\cal K} \left| \phi \right>$ with the definition
\begin{eqnarray}
\label{LUPR}
{\cal K} = 
\frac{1}{ \sqrt{\left< \left( \Psi | \right| K  \right)
                    \! \left( K | \!\left| \Psi \right) \right>} }
\left| K \right) \! \left( K \right| .
\end{eqnarray}
When a partial measurement is made on the system in the state
$\left| \left| \Psi \right) \right> $ of  (\ref{fullwf}),
it turns into
\begin{eqnarray}
\label{AFTM}
\left| \left| \Psi' \right) \right> =
\frac{1}{ \sqrt{\left< \left( \Psi | \right| K  \right)
                    \! \left( K | \!\left| \Psi \right) \right>} }
 \left| K \right) \! \left( K | \!\left| \Psi \right) \right> 
\end{eqnarray}
The partial matrix element $\left( K \right| \left. \!\!\left| \Psi \right) \right> $
is calculated as
\begin{eqnarray}
&&\!\!\!\!\!\!\!\!\!\!\!\!\!\!\!\!
\left( K \right| \left. \!\!\left| \Psi \right) \right> = 
\left< 0 \right| \!\left( K \right| \left. \!\!\left| \Psi \right) \right> \left| 0 \right>  + 
\left< 1 \right| \!\left( K \right| \left. \!\!\left| \Psi \right) \right> \left| 1 \right> 
\nonumber \\
&&\ \ 
=
(\kappa^{(0)*} \psi^{(0)}_0+ \kappa^{(1)*}\psi^{(1)}_0) \left| 0 \right>  + 
(\kappa^{(0)*} \psi^{(0)}_1+ \kappa^{(1)*}\psi^{(1)}_1) \left| 1 \right> .
\end{eqnarray}
In the state $\left| \left| \Psi' \right) \right>$, the absolute value squared of
the coefficient in front of 
the state $\left| K \right) \! \left| j \right>$ gives the probability of the agent's  
action $\left< j \right>$ under the condition of the occurrence of the
mixed event $\left |K\right)$, which we denote as $P^K_j$.  
We have
\begin{eqnarray}
\label{PKQN}
P^K_j =
\frac{ \left|{ \left< j \right| \left( K \right| \left. \!\!\left| \Psi \right) \right>  }\right|^2}
   {\left|{ \left< 0 \right| \left( K \right| \left. \!\!\left| \Psi \right) \right>  }\right|^2
  +\left|{ \left< 1 \right| \left( K \right| \left. \!\!\left| \Psi \right) \right>  }\right|^2} ,
\end{eqnarray}
which leads to the quantum description of the conditional probability $P^K_j$ 
in the form 
\begin{eqnarray}
\label{QJP}
P^K_j= \frac
{q^{(0)} p^{(0)}_j + q^{(1)} p^{(1)}_j 
  + 2 \sqrt{ q^{(0)} q^{(1)} p^{(0)}_j p^{(1)}_j }\cos\theta_j}
{1+ 2 \sqrt{ q^{(0)} q^{(1)} p^{(0)}_0 p^{(1)}_0 }\cos\theta_0
   + 2 \sqrt{ q^{(0)} q^{(1)} p^{(0)}_1 p^{(1)}_1 }\cos\theta_1} .
\end{eqnarray}
Here, the phase $\theta_j$ is defined through
\begin{eqnarray}
\kappa^{(0)}\kappa^{(1)*} \psi^{(0)*}_j\psi^{(1)}_j 
= e^{i\theta_j}  \sqrt{ q^{(0)} q^{(1)} p^{(0)}_j p^{(1)}_j } .
\end{eqnarray}
We clearly see that the quantum description has two extra 
phase parameters $\theta_1$ and $\theta_2$, on top of classical 
probabilities $q^{(k)}$ and $p^{(k)}_j$. 
These extra parameters might be thought of as representing
internal psychological traits of the agents
that admix the consideration of geometric average for intermediate event
to the conventional arithmetic average. 
The $\theta$-dependent term in the enumerator represents the quantum interference
of probabilities, while the ones in the denominator are the terms coming from 
``wave function renormalization'' whose existence has been first noted in \cite{KH09}. 
The general and explicit expressions of quantum conditional probability, (\ref{PKQN}) 
and (\ref{QJP}) are the main result of the formal side of this work. 
%
The probability $P_j^{K}$ is reduced to the one given by the classical description
\begin{eqnarray}
\label{CLP}
P_j^{K} [cl] =  q^{(0)} p^{(0)}_j + q^{(1)} p^{(1)}_j ,
\end{eqnarray}
when these phase parameters have particular values 
$\theta_1 = \theta_2 = \pi$.
It is obvious from (\ref{CLP}), that, in classical description, 
the conditional probability $P^K_j$ necessarily falls   
between $p^{(0)}_j$ and $p^{(1)}_j$ because $q^{(0)}$ and $q^{(1)}$ are 
positive numbers adding up to the unity, namely  
\begin{eqnarray}
\label{cbound}
\min ( p^{(0)}_j ,  p^{(1)}_j ) \le P^K_j \le
\max ( p^{(0)}_j ,  p^{(1)}_j )
\qquad({\rm Classical}).
\end{eqnarray}
This fact, that we should expect intermediate
probability for intermediate event, is called {\it sure-thing principle}
in psychological context.  
%
%
The necessity of the denominator in (\ref{QJP}), for general value of phases,
is related to the fact that the absolute values of 
two matrix elements $\left< 0 \right| \!\left( K \right| \left. \!\left| \Psi \right) \right>$ 
and $\left< 1 \right| \!\left( K \right| \left. \!\left| \Psi \right) \right>$ not summing
up to one.
This occurs because the conditional event $\left| K \right)$ alone does not
exhaust the Hilbert space of preconditions, but has to be
supplemented by the complementary state $\left| {\bar K} \right)$ defined
as
\begin{eqnarray}
\left| {\bar K} \right) = \kappa^{(1)*} \left| 0 \right)- \kappa^{(0)*} \left| 1 \right).
\end{eqnarray}
With this state, which is orthogonal to $\left| K \right)$, we have the completeness
\begin{eqnarray}
\left| K \right) \left( K \right| + \left| {\bar K} \right) \left( {\bar K} \right| = 1,
\end{eqnarray}
and conditional probabilities 
$\left< j \right| \!\left( K \right| \left. \!\left| \Psi \right) \right>$ 
and $\left< j \right| \!\left( {\bar K} \right. \left| \left| \Psi \right) \right>$, 
with $j=0, 1$, do add up to unity.
%

We note that our treatment of intermediate precondition, (\ref{AFTM}), is essentially identical, in the language of quantum measurement theory, to the L{\"u}ders' projection postulate \cite{LU51} applied to a partial one-body measurement of a two-body system.  If we consider repeated measurements with the same intermediate state $|K)$, it is reduced to the standard von Neumann's projection postulate \cite{NE55} with the pure state density matrix $\rho = |K)(K|$.  If we further replace the pure state by a mixed state made up of states with incoherent random phases for $\kappa^{(0)}$ and $\kappa^{(1)}$, the interference effects cancel out among themselves \cite{GI84}.  The von Neumann postulate then yields the classical result, (\ref{CLP}).

%
\section{Quantum bound of conditional probability}
%
%
%

Now, going back to the general expression (\ref{QJP}), we consider the
maximum and minimum for the quantum conditional probability.
We first define the complementary value of $j$ as ${\bar j} = 1 - j$. 
Since the denominator is always positive, $\theta_{\bar j}=\pi$ will
minimize it, and thus maximize  $P^K_j$.  We then have
\begin{eqnarray}
\label{QJP1}
P^K_j
&&\!\!\!\!\!\!\!\!\!\!
= \frac
{q^{(0)} p^{(0)}_j + q^{(1)} p^{(1)}_j 
  + 2 \sqrt{ q^{(0)} q^{(1)} p^{(0)}_j p^{(1)}_j }\cos\theta_j}
{1- 2 \sqrt{ q^{(0)} q^{(1)} p^{(0)}_{\bar j} p^{(1)}_{\bar j} }
   + 2 \sqrt{ q^{(0)} q^{(1)} p^{(0)}_j p^{(1)}_j }\cos\theta_j} 
\nonumber \\   
&&\!\!\!\!\!\!\!\!\!\!
=1-
 \frac
{\left( \sqrt{ q^{(0)} p^{(0)}_{\bar j} } - \sqrt{ q^{(1)} p^{(1)}_{\bar j} } \right)^2}
{1- 2 \sqrt{ q^{(0)} q^{(1)} p^{(0)}_{\bar j} p^{(1)}_{\bar j} }
   + 2 \sqrt{ q^{(0)} q^{(1)} p^{(0)}_j p^{(1)}_j }\cos\theta_j}
\end{eqnarray}
which takes the maximum value with the choice $\theta_{j}=0$
Similarly, minimum value of $P_j^K$ is shown to be obtained at
$\theta_j=\pi$ and $\theta_{\bar j} = 0$. We have
\begin{eqnarray}
\label{INEQ}
\frac{\left( \sqrt{q^{(0)} p^{(0)}_j} - \sqrt{q^{(1)} p^{(1)}_j} \right)^2}{1-f_j}  
\le P^K_j \le 
\frac{\left( \sqrt{q^{(0)} p^{(0)}_j} + \sqrt{q^{(1)} p^{(1)}_j} \right)^2}{1+f_j} 
\end{eqnarray}
%
where the correction term $f_j$ is give by
\begin{eqnarray}
f_j = 2\sqrt{ q^{(0)} q^{(1)}}
\left( \sqrt{p^{(0)}_j p^{(1)}_j }-\sqrt{p^{(0)}_{\bar j} p^{(1)}_{\bar j} } \right) .
\end{eqnarray}
%
%

For the sake of arguments, let us for a moment
limit ourselves to consider only those cases
in which $f_j$ can be neglected.  This occurs exactly when we have
%
$
p^{(0)}_0 = p^{(1)}_1 
$
%
which also imply $p^{(0)}_1 = p^{(1)}_0$ naturally.
We then only need to look at the enumerators of (\ref{INEQ})
to identify the two bounds;
\begin{eqnarray}
\frac{(2q^{(0)} p^{(0)}_j) + (2q^{(1)} p^{(1)}_j) }{2}
\pm \sqrt{ (2q^{(0)}p^{(0)}_j)  (2q^{(1)} p^{(1)}_j) }.
\end{eqnarray}
We can now express the inequalities as
{\it quantum conditional probability of intermediate event
is bounded by the sum and difference
of weighted arithmetic and geometric means of probabilities of two events}.

Going back to the general case, we consider the situation in which 
even the probability of the condition, $q^{(k)}$ is not known.
We vary $q^{(0)}$  in (\ref{INEQ}) to maximize both bounds, and obtain
\begin{eqnarray}
\label{UNK1}
P^K_j(q_{max}^{(0)}) = 1,
\quad
P^K_j(q_{min}^{(0)}) = 0, 
\end{eqnarray}
with
\begin{eqnarray}
\label{UNK2}
q_{max}^{(0)} = \frac{ \left(\sqrt{p_j^{(0)}} - \sqrt{p_j^{(1)}} X_j \right)^2  }
{ \left(\sqrt{p_j^{(0)}} - \sqrt{p_j^{(1)}} X_j \right)^2 
   + \left(\sqrt{p_j^{(1)}} - \sqrt{p_j^{(0)}} X_j \right)^2  }
\nonumber \\
q_{min}^{(0)} = \frac{ \left(\sqrt{p_j^{(1)}} - \sqrt{p_j^{(0)}} X_j \right)^2  }
{ \left(\sqrt{p_j^{(0)}} - \sqrt{p_j^{(1)}} X_j \right)^2 
   + \left(\sqrt{p_j^{(1)}} - \sqrt{p_j^{(0)}} X_j \right)^2  }
\end{eqnarray}
where
\begin{eqnarray}
X_j = \sqrt{p_j^{(0)}p_j^{(1)}}-\sqrt{(1-p_j^{(0)})(1-p_j^{(1)})}. 
\end{eqnarray}
We therefore obtain the quantum counterpart of sure-thing principle in the form of 
a trivial relation
\begin{eqnarray}
\label{QSP}
0 \le P^K_j \le 1 
\qquad({\rm Quantum}),
\end{eqnarray}
meaning that we should expect any value for conditional probability under unknown
combination of preconditions.
This is to be compared to the classical bound of sure-thing principle, (\ref{cbound}).
%

\section{Numerical examples}

From practical point of view, in analyzing the results of psychological experiments,
the main results of the previous section can be summarized as follows.

\bigskip
(i) Known unknown: 
When the probability of conditional event $q^{(k)}$ is known, the conditional
probability for the event $K$, which is made up of probabilistically known 
combination of events $\left|0\right)$ and $\left|1\right)$, can take any value constrained by equation
(\ref{INEQ}) with proper choice of two phase parameters $\theta_1$ and $\theta_2$. 

\bigskip
(ii) Unknown unknown:
When even the probability of conditional event $q^{(k)}$ is unknown, $P^{K}$, the conditional
probability for the event $\left|K\right)$,  can take any value between zero and one.

\bigskip
In order to scrutinize the predictive power of quantum decision theory, therefore, 
we need to have more than two experimental numbers for probability $P^{K}$ with 
different inputs $p^{(0)}_j$ 
and $p^{(1)}_j$, for a given ``mind set'' represented by  a set $(\theta_1, \theta_2)$ 
and a known $q^{(0)}=1-q^{(1)}$.  Conversely, any two sets of experimental numbers
($p^{(0)}_j$, $p^{(1)}_j$, $P^{K}_j$), which are 
within the bound of (\ref{INEQ}), can be fitted with some combination of values 
for $\theta_1$ 
and $\theta_2$.
For a situation in which even the values of $q^{(0)}$ is unknown  (but known to stay
in a fixed value), the number of experimental data to be tested has to be more than three.
  
%
%
%
\begin{figure}
\center{
\includegraphics[width=5cm]{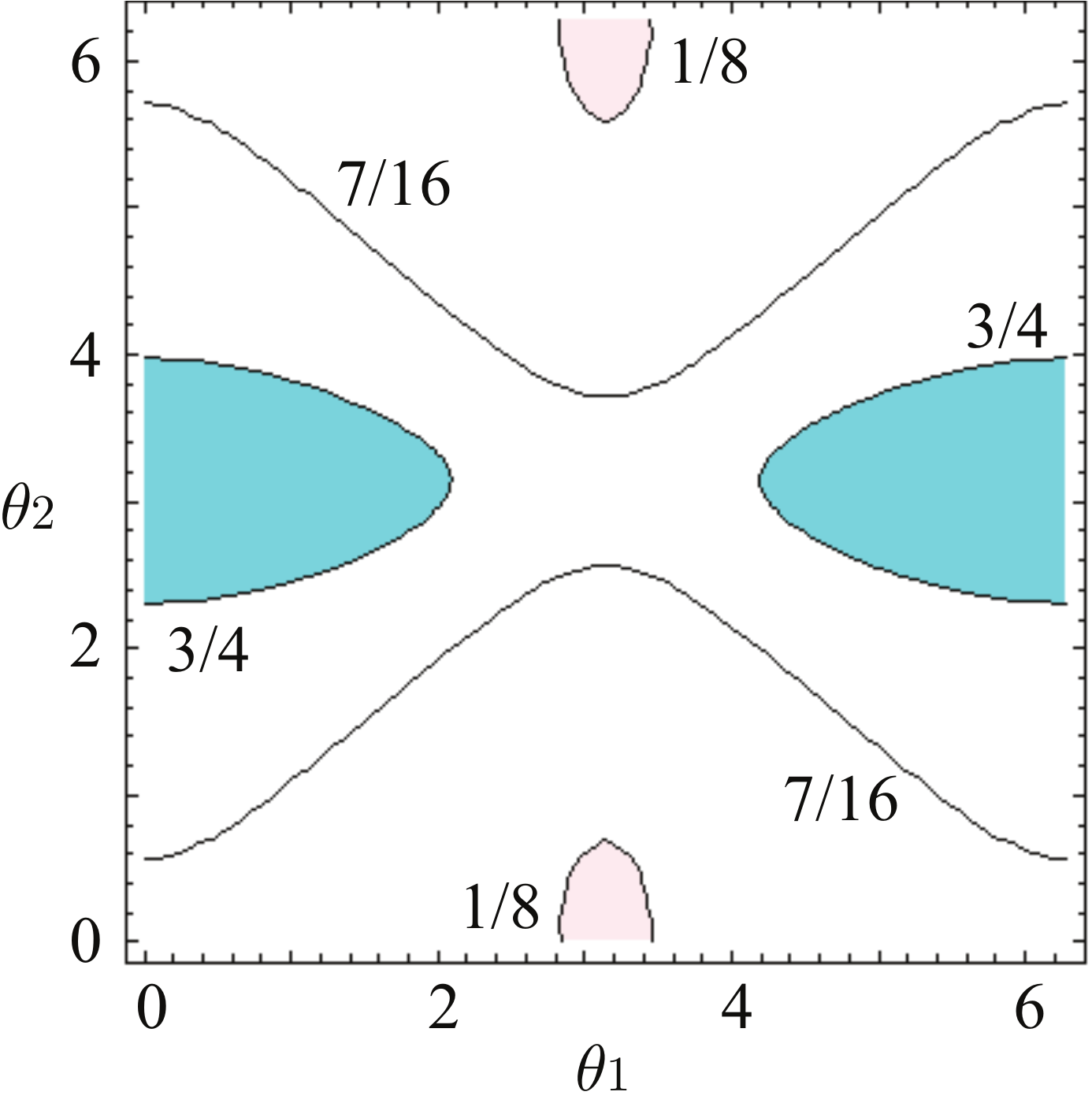}\quad
\includegraphics[width=5cm]{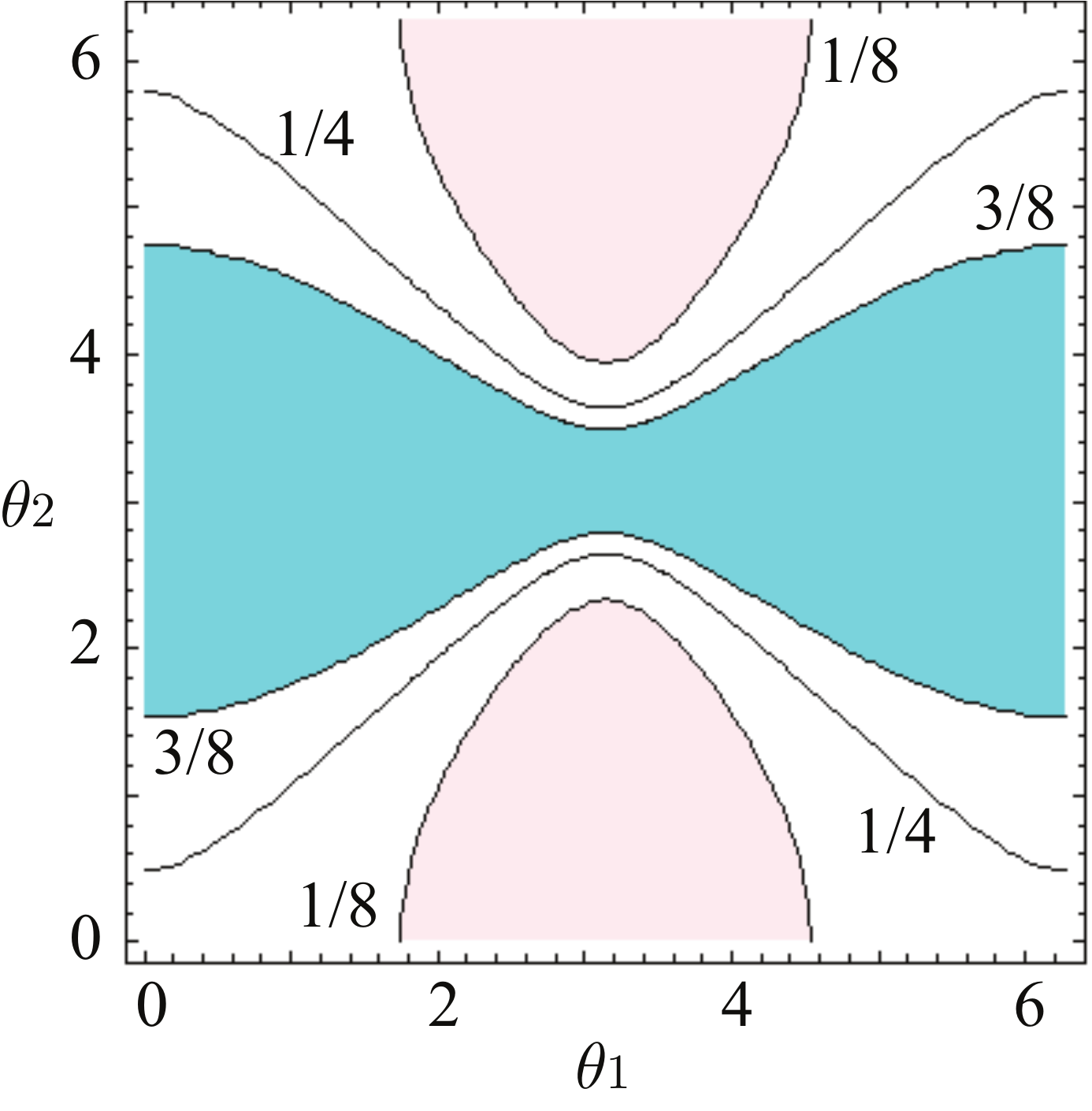}
}
\label{fig1}
\caption
{
Contour graphs representing $P^K$ as a function of quantum phases $\theta_1$ 
and $\theta_2$.
The white region represent the classical bound, inside of which the curve 
representing the classical prediction $P^{K}_0[cl]$ lies.  The region 
with light shade (or pink) and with dark shade (or cyan) is where $P^{K}_0$ is 
below and above classical bound, respectively.  The left side is the result of inputs
$p^{(0)}=\frac{6}{8}$ and $p^{(1)}=\frac{1}{8}$, while the right side, of
$p^{(0)}=\frac{3}{8}$ and $p^{(1)}=\frac{1}{8}$.
}
\end{figure}

We first look at a fictitious two examples to get an idea how quantum theory works to provide numbers not attainable by classical probabilities.
In Figure 1, we show contour graphs for  $P^{K}=P^{K}_0$ 
as a function of  $(\theta_1, \theta_2)$ with a
given $(p^{(0)}_0, p^{(1)}_0) = (p^{(0)}, p^{(1)})$ .  
We immediately notice the mirror symmetry along the axes $\theta_1=\pi$
and $\theta_2=\pi$, which is in fact evident from (\ref{PKQN}).
The probability of condition is assumed to take 
$q^{(0)}=q^{(1)}=\frac{1}{2}$.  In the graph in the left, we set  
$( p^{(0)}, p^{(1)} )$ $= (\frac{1}{8}, \frac{6}{8})$, while in the graph in the right,
it is $( p^{(0)}, p^{(1)})$ $= (\frac{1}{8}, \frac{3}{8})$.  The white region is the
value within classical bound of sure-thing principle, and the curved line in the white region 
represents the classical prediction $P^{K}[cl]$.  The region with light shade (or pink) 
is where $P^{K}$ is below $\min(p^{(0)}, p^{(1)})$ and the region with dark
shade (or cyan) is where $P^{K}$ is above $\max(p^{(0)}, p^{(1)})$.
When $p^{(0)}$ and  $p^{(1)}$ are closer, $P^{K}=0$ is more likely to
get beyond classical sure-thing principle limit, as predicted by the quantum formula 
(\ref{PKQN}).  In fact, for the case of fixed $q^{(0)}$, it is naturally not so much the violation of 
sure-thing principle, but the deviation form the classical value itself,
that calls for the quantum explanation. 
We now go on to examine the numbers from real world experiments performed up to now, 
which are summarized in Table 1.  In these experiments, each participants is asked 
to choose from ``bad' or ``good'' behavior toward a fictitious opponent under certain condition
on the knowledge of the choice of opponents (specified by $k$), and the percentages
of people making ``bad'' and ``good'' choices are recorded as $p^{(k)}_0=p^{(k)}$ 
and $p^{(k)}_1=1-p^{(k)}$.  
Participants were told that they and their opponents were under prisoner's dilemma-type
situation, where one's ``bad'' choice would be rewarded in the expense of his opponent,
but he and his opponent would both benefit by both making ``good'' choice together, 
compared to both making ``bad'' choices.  Experiments were done with real financial
reward at stake.
Experiments are done under three conditions.
In one of them ($k=0$), they are told that the opponent have chosen the ``bad'' strategy, 
and in another ($k=1$),  they are told that the other have gone ``good''.   
In the third situation, they are told that the choice of the other is unknown ($K$),
for which case, the probability of participants' choosing ``good'' and ``bad'' behaviors are
written as $P^K_0=P^K$ and $P^K_1=1-P^K$.

For all experiments, it is difficult to estimate the probabilities $q^{(k)}$, 
since this number is to represent ``unknown'' condition,  not a testable quantity.  
Here, we simply assume, as a working hypothesis,  that
they are given by $q^{(0)}=q^{(1)}=\frac{1}{2}$.
In all examples in Table 1, the numbers for $P^K$ are not only far from
the classical prediction $P^K[cl]$ given by (\ref{CLP}),  but also are found to 
fall below $\min(p^{(0)}, p^{(1)})$, which is the limit set by sure-thing principle.
Namely, sure-thing principle is broken in all four experiments.
It is also clear that they are well within the quantum bounds, (\ref{INEQ})
whose lower and upper bounds are written as $P^K_{min}$ and $P^K_{max}$ 
in the Table 1.
%
%
\begin{table}[ht]
\center{
\begin{tabular}{c|ccc|c|cc}
\hline
Authors / Year & $p^{(0)}$ & $p^{(1)}$ & $P^{K}_{exp}$ & $P^{K}[cl]$ &
 $P^{K}_{min}$ & $P^{K}_{max}$ \\ 
\hline
Shafir \&Tversky / 1992 \cite{ST92} & 0.97 & 0.84  & 0.63 & 0.91 & 0.02 & 0.98 \\ 
Croson / 1999 \cite{CR99} & 0.67  & 0.32 & 0.30 & 0.45 & 0.03 & 0.97 \\ 
Li \& Taplin / 2002 \cite{LT02} & 0.83  & 0.66 & 0.60 & 0.75 & 0.01 & 0.99 \\ 
Busemeyer {\it et al.} / 2006 \cite{BM06} & 0.91  & 0.84 & 0.66 & 0.88 & 0.00 & 1.00 \\ 
\hline 
\end{tabular}
\caption{Summary of experimental numbers of the conditional probabilities, 
$p^X=p^X_0 = 1 -p^X_1$,  
representing the percentage of 
participants choosing  ``bad'' move under different conditions signified by
$X=(0), (1)$ and  $K$.  First three columns 
are the experimental data, the fourth, the classical prediction for $p^K$, and the last two, the quantum lower and upper bounds.}
}
\end{table}
%

%
%
\begin{figure}[ht]
\center{
\includegraphics[width=4.2cm]{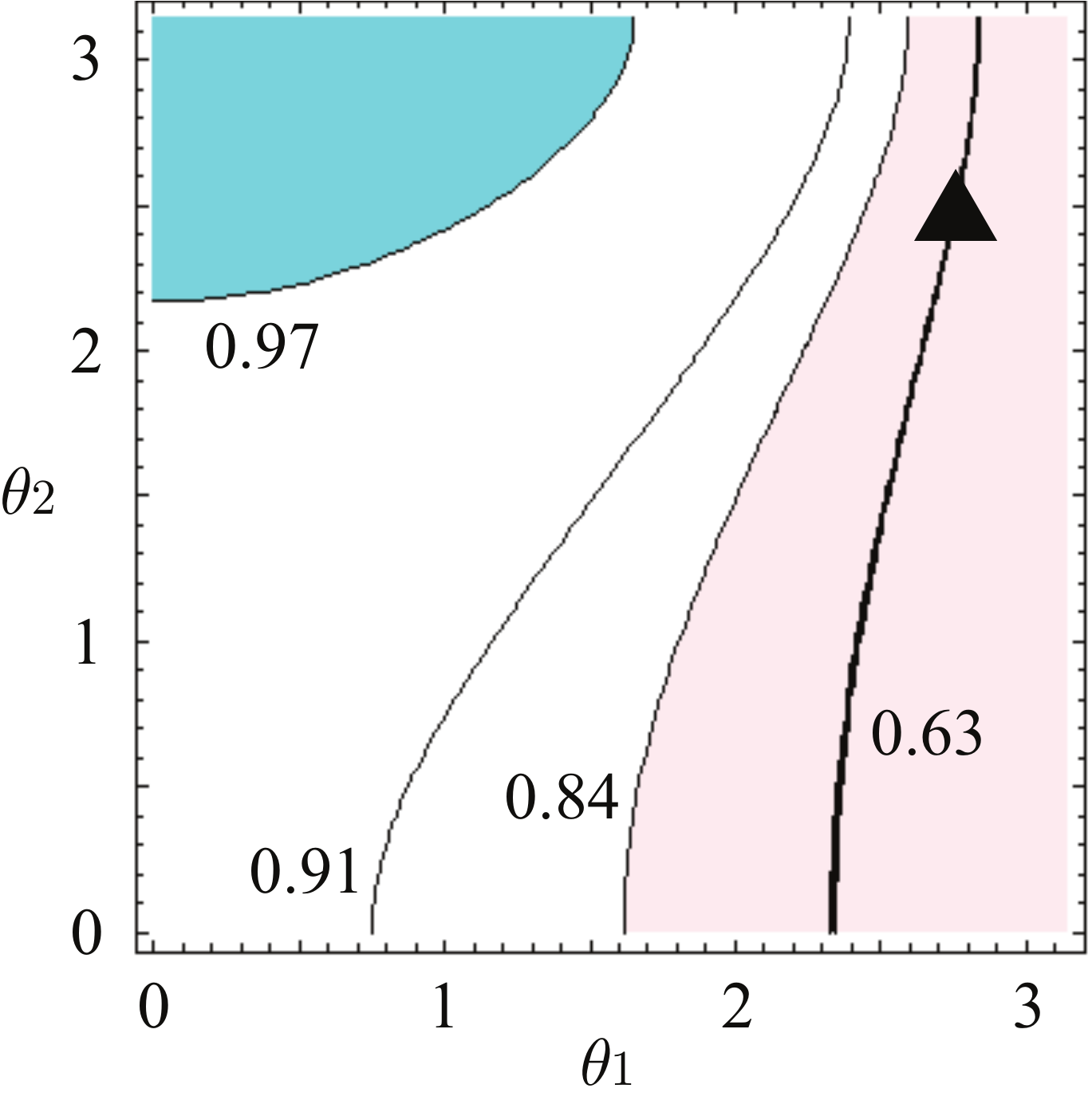}\quad
\includegraphics[width=4.2cm]{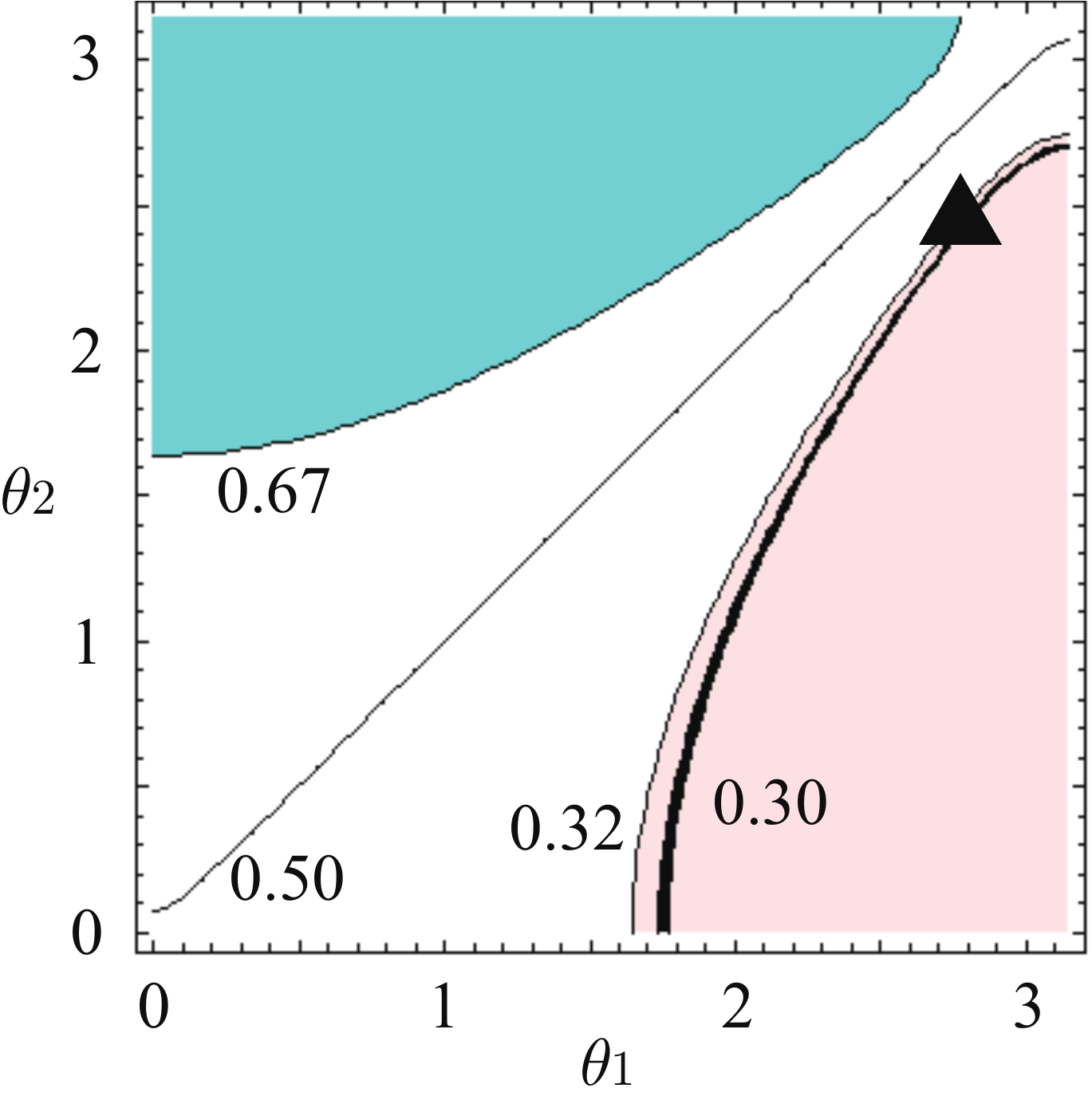}\\ \bigskip
\includegraphics[width=4.2cm]{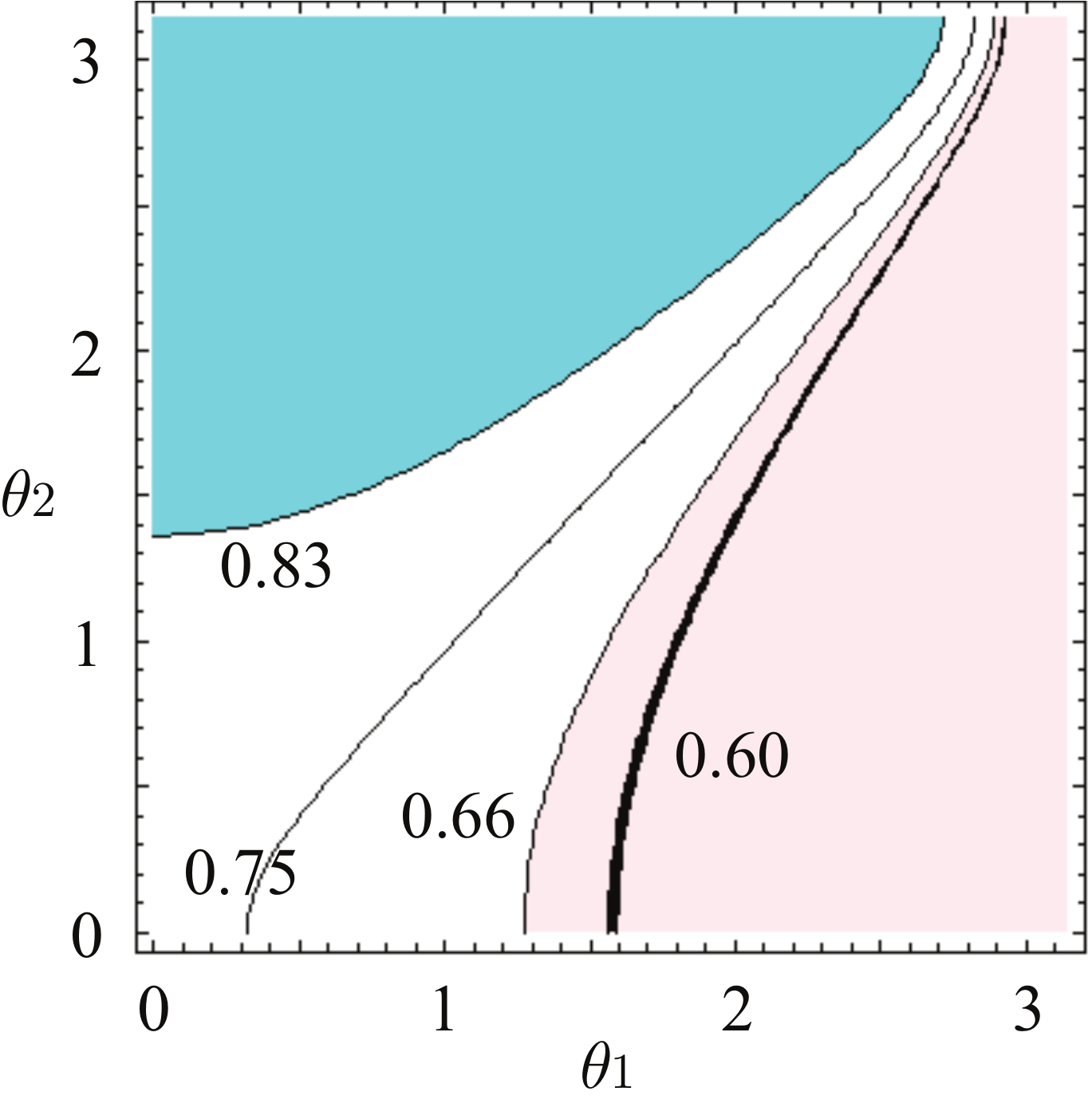}\quad
\includegraphics[width=4.2cm]{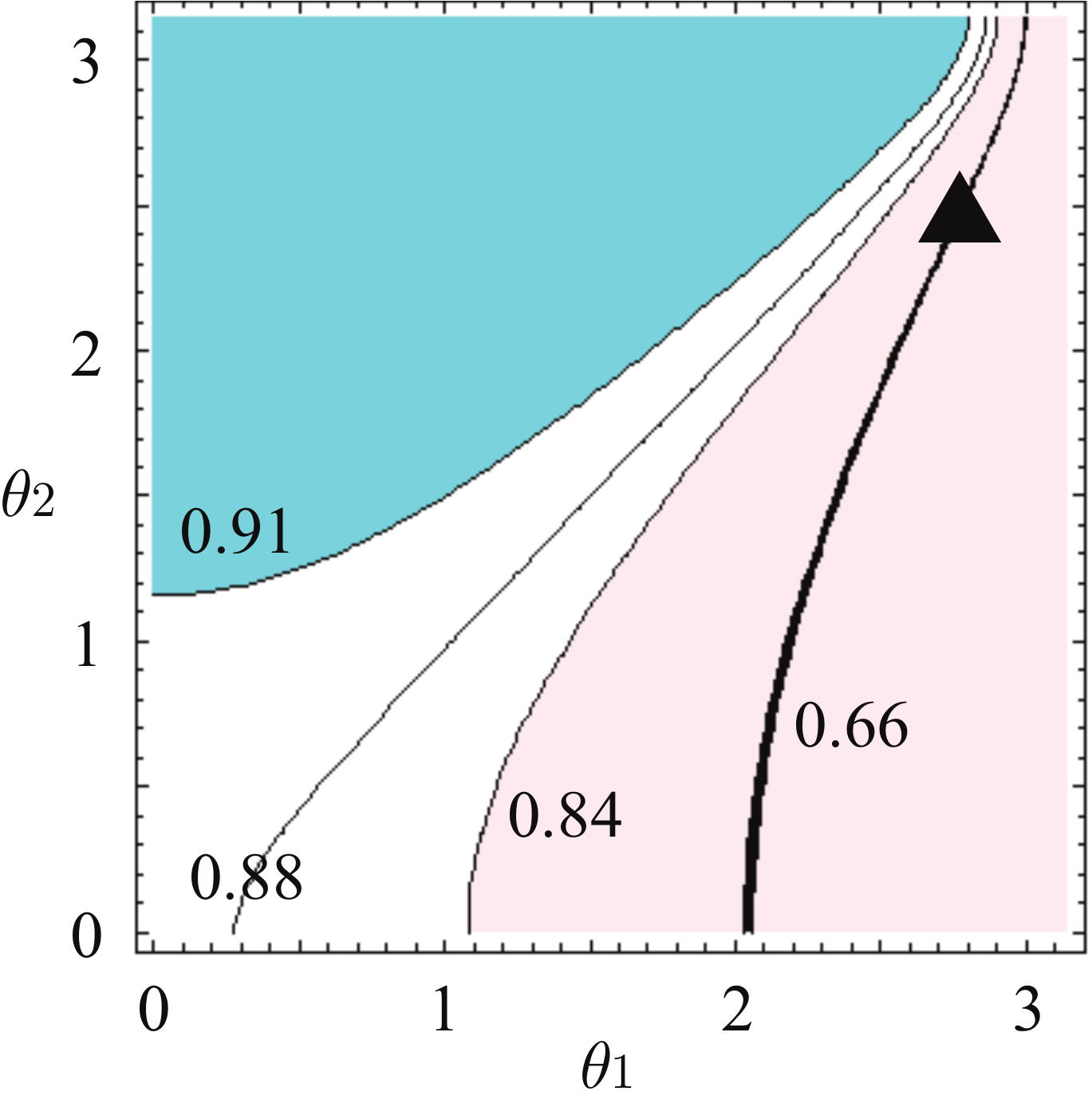}
}
\label{fig2}
\caption
{
Contour graphs representing $P^K$ as a function of quantum phases $\theta_1$ 
and $\theta_2$.  Values of $p^{(0)}$ and $p^{(1)}$ are 
taken from experiments found in Table 1.
The white region represent the classical bound, inside of which the curve 
representing the classical prediction $P^{K}_0[cl]$ lies.  The region 
with light shade (or pink) and with dark shade (or cyan) is where $P^{K}_0$ is 
below and above classical bound, respectively.  The solid line is the trajectory of
$(\theta_1, \theta_2)$ giving the experimental $P^{K}$.  The triangles in three of the graphs 
points to the common value of $(\theta_1, \theta_2)$ where their solid lines cross each other. 
}
\end{figure}

In Figure 2, we show contour plot of $P^K$ as a function of two quantum phase angles
$(\theta_1, \theta_2)$. 
As before, the white region is the value of $P^K$ 
within classical bound of sure-thing principle, 
and the curved line in the white region represents the classical prediction $P^{K}[cl]$.  
The regions with light shade (or pink)  and dark shade (or cyan) are 
where $P^K$ is below or above the classical sure-thing principle limit, and solid line,
falling within light-shaded (pink) region in all four examples, 
represents the set of phases $(\theta_1, \theta_2)$ that gives
the experimentally observed $P^K$ values. 

Broadly speaking, all graphs share similar appearance.  But upon close examination
by superimposing them, we observe the following two points:

\bigbreak
(i) The result of Li and Taplin \cite{LT02} is an ``odd man out'', and 
the experimental solid line of
the graph of Li and Taplin do not share any point with the lines from the
other three experiments.

\bigbreak
(ii) The results of Shafir and Tversky \cite{ST92}, Croson \cite{CR99}, and 
Busemeyer {\it et. al.} \cite{BM06} give a consistent single choice of
the phase   $(\theta_1, \theta_2) \approx (0.27, 0.25)$, marked by the solid triangle,
at which point the experimental solid lines of three graphs cross each other.

\bigbreak
A very optimistic interpretation is that in the three ``consistent'' experiments,
groups of participants share common psychological traits that is indeed
successfully described by quantum decision theory with a single set of quantum phases
$(\theta_1, \theta_2)$.
And in  Li and Taplin experiment, the psychological makeup of the
participants was markedly different from all other
experiments.  This experiment is presumably described by quantum theory with 
some other value of phases, which should lie somewhere on the solid line.
A somberer view is that, we still lack both sufficient number and sufficient accuracy
in the experimental data, and it is too early to call either success or the failure of
quantum description of these psychological conditional probabilities.

In either way, through these numerical examples, 
we now have a clearer view on how to sort out the experimental numbers.
With the accumulation of more experimental data, 
preferably with finer control of conditions, we
may eventually hope to judge phenomenological applicability and predictive power 
of quantum decision theory.

\section{Prospects}
In this work, 
we have identified the minimal additional element of quantum theory 
{\it viz {\`a} viz} classical theory of  conditional decision probability 
in two-by-two settings.
It is straightforward to extend the treatment here to the case of more than
two conditional events, and also to more than two choices for the agent.
Suppose there are $M$ conditional events, to each of which there are $N$ choices.
The number of relative phase appearing in the expression of conditional probability 
is the ${}_N C_2$, and there should be $M$ of these quantities.  Thus the number
of newly introduced parameters in the quantum description is $\frac{MN(N-1)}{2}$.

%
%
The success of quantum decision theory, if it is eventually obtained with
more data, can mean one of two things.  
It can simply represent successful phenomenology with 
sufficient number of parameters and sufficient flexibility in formulation that
can effectively simulate a more involved classical decision theory with sub-divided 
psychological conditions and cases.  It can also mean a genuine quantum nature 
of some elements in psychological process in decision making.
A parallel could be drawn from the quantum game theory \cite{ME99, EW99}, 
which is shown to be applicable, on one hand, to non-quantum settings due to 
its effective inclusion of altruistic game strategy  \cite{CH03}, and, on the other hand,
is shown to include truly quantum effects that come from
quantum interference  \cite{CT06}.
To rigorously test the existence or the absence of genuinely quantum effect,
we might need to consider decision making experiment with incomplete information,
analogous to the Harsanyi type quantum game \cite{CI08} in which breaking of 
the Bell inequality \cite{BE64} can be directly tested.  
 

\bigskip
{\noindent}{\bf Acknowledgments} \\
This work has been partially supported by 
the Grant-in-Aid for Scientific Research of  Ministry of Education, 
Culture, Sports, Science and Technology, Japan
under the Grant Numbers  21540402 and 20700236.

\end{document}